# Non-volatile Spin Switch for Boolean and Non-Boolean Logic

Supriyo Datta
School of ECE, Purdue University, IN 47907, USA
datta@purdue.edu

Sayeef Salahuddin
Department of EECS, UC Berkeley, CA 94720, USA

Behtash Behin-Aein
GLOBALFOUNDRIES Inc., Sunnyvale, CA 94085, USA

*Abstract* – We show that the established physics of spin valves together with the recently discovered giant spin-Hall effect could be used to construct *Read* and *Write* units that can be integrated into a single spin switch with input-output isolation, gain and fan-out similar to CMOS inverters, but with the information stored in nanomagnets making it non-volatile. Such spin switches could be interconnected, with no external amplification, just with passive circuit elements, to perform logic operations. Moreover, since the digitization and storage occurs naturally in the magnets, the voltages can be used to implement analog "weighting" for non-Boolean logic.

1. Introduction
2. *Read* unit
3. *Write* unit
4. Concatenable spin switch
5. Ring oscillator
6. Reconfigurable comparator
7. Weighted interconnections
8. Summary
    Appendix A: Modeling a spin switch
    Appendix B: Modeling spin switch networks

## 1. INTRODUCTION

The advances of the last two decades have effectively integrated two distinct fields of research, spintronics and nanomagnetics, into one. Spin valves (SVs) and magnetic tunnel junctions (MTJ's) are now used both to read from and to write to nanomagnets. This has transformed the field of memory devices and it is natural to ask whether logic devices too can be designed to take advantage of these seminal advances [1].

Performing logic operations is conceptually straightforward if external amplifiers or clocking circuitry are used to interface the output from a *Read* unit to the input of a *Write* unit. This paper, however, is about autonomous or *self-contained logic (SCL)* units that can be interconnected to perform logic operations just with passive circuit elements and a power supply, without external amplifiers or clocks. This requires an SCL unit to exhibit internal gain and directivity, attributes that are well-known for transistors, but not for magnets. Of course one may still want to use clocks to regulate the flow of data as one does with transistor circuits, but it is not a necessary component for fundamental logic operations.

In earlier work on all-spin logic (ASL) we discussed the possibility of implementing such SCL units using bistable nanomagnets as digital spin capacitors that act as threshold elements interacting via spin currents $\vec{I}_s$

$$\hat{m} \to \vec{I}_s \xrightarrow{Transmit} \vec{I}_s \to \hat{m}'$$

in a non-reciprocal manner [2] based on the phenomenon of non-local spin transfer torque (NLSTT) [3,4] (Fig.1a). The purpose of this paper is to draw attention to the possibility of a scheme (Fig.1b) which also exploits the digital nature of magnets like ASL, but the communication occurs through charge currents I instead of spin currents (Fig.1b):

$$\hat{m} \to I \xrightarrow{Transmit} I \to \hat{m}'$$

Although charge currents carry less information than spin currents, they should be adequate for many applications while making the communication more robust and long range.

In our proposed scheme the *Read* unit (see Fig.1b) is based on magnetic tunnel junctions (MTJ) or SVs while the *Write* unit is based on the giant spin Hall effect (GSHE). In principle the *Write* unit could be implemented using an ordinary spin transfer torque (STT) device, but the charge current I would then be larger than the spin current $I_s$ needed to switch the magnet. By contrast, recent experiments [5] have clearly shown that the GSHE allows the use of a current I considerably less than $I_s$. This makes it relatively straightforward to design our spin switch with the internal gain and directivity characteristic of SCL units that can be used to build circuits without external amplifiers or clocks. We leave it to future work to determine whether other writing mechanisms can be used to achieve similar gain and directivity.

We will show that like ASL, the spin switch meets the five basic tenets required of a digital logic device [6]: It is inherently *non-linear* due to the bistable nature of magnets and the output is electrically isolated from the input with *negligible feedback*. It is *concatenable*, that is, the output of one unit can drive the input of the next, can be designed to provide *gain* based on present-day technology and can be used to construct a *complete set* of Boolean logic gates.

One disadvantage of the structure in Fig.1b is that it draws current continuously which could be reduced a little if the writing mechanism were voltage-driven. Even with the current-driven mechanism we are discussing, it may be preferable to put the resistor "r" in series with a capacitor which reduces the standby current. Appendix B describes a simulated operation of a majority gate with fan-out including capacitors in series with the "r".

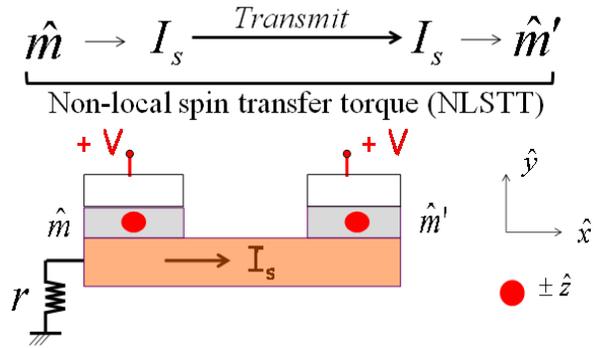

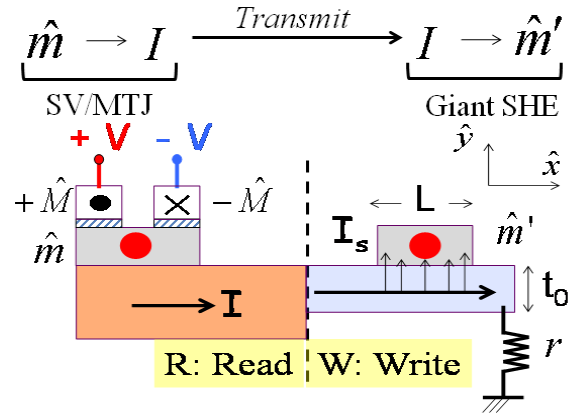

Figure 1: (a) All-spin logic (ASL) proposes to store information in nanomagnets which communicate through spin currents, based on the physics of non-local spin transfer torque (NLSTT). (b) This paper proposes an alternative scheme: Information stored in nanomagnets is communicated through charge currents instead of spin currents. The *Read* unit is based on the established physics of spin valves (SV) / magnetic tunnel junctions (MTJ) while the *Write* unit is based on the giant spin-Hall effect (SHE). This approach is shown to allow *self-contained logic* implementation, like ASL.

We estimate the energy-delay product of our proposed spin switch to be inferior to complementary metal oxide semiconductor (CMOS) inverters, but this could change with further advancements in spintronic switching of magnets and requires a careful discussion beyond the scope of this paper. On the other hand, unlike CMOS devices the information is stored in nanomagnets and not in the voltage, making it non-volatile. It also naturally allows for non-Boolean operation as we will illustrate with a simple example of a reconfigurable comparator (Fig. 5), an ability not ordinarily available in a CMOS implementation.

Let us start with the basic physics underlying the *Read* and *Write* units that comprise the spin switch and then show how these units can be integrated to provide a concatenable device that we call a spin switch.

## 2. *READ* UNIT

The read device consists of two nanopillars on top of the input magnet $\hat{m}$ (Fig.1b), each representing an spin valve (SV) or an magnetic tunnel junction (MTJ) whose conductance is determined by the orientation of the input magnet relative to the fixed magnets $\hat{M}$ and $-\hat{M}$.

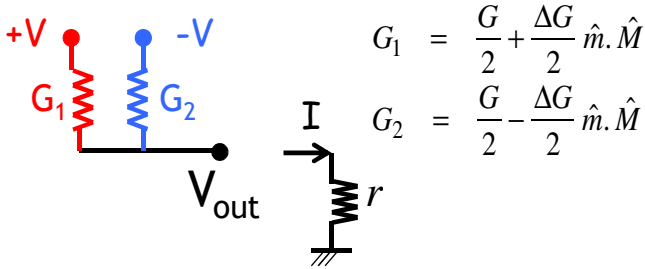

$$G_1 = \frac{G}{2} + \frac{\Delta G}{2} \hat{m}.\hat{M}$$
$$G_2 = \frac{G}{2} - \frac{\Delta G}{2} \hat{m}.\hat{M}$$

Figure 2 – Equivalent circuit used to analyze the *Read* unit in Fig.1b.

This structure can be modeled with the equivalent circuit shown in Fig.2, where G and $\Delta G$ represent the sum and difference respectively of the parallel ($G_P$) and anti-parallel ($G_{AP}$) conductances of the SV or MTJ. This circuit leads straightforwardly to the following expression for the open circuit voltage and the current:

$$V_{out} = \frac{V \Delta G}{G} \hat{m}.\hat{M}$$
$$I = \frac{V_{out}}{r + 1/G} \quad (1)$$

$V_{out}$ is thus proportional to the component of the input magnetization $\hat{m}$ along a fixed direction $\hat{M}$ determined by the fixed magnets. Note that $\Delta G = G_P - G_{AP}$ and $G = G_P + G_{AP}$. Assuming a TMR of 135% and resistance-area product of $A/G_P = 4.3$ ohm-$\mu m^2$ [7] we estimate a polarization and a conductance of

$$P \equiv \Delta G / G = TMR/(TMR+2) = 0.4$$
$$G \sim (1.1 \, K-ohm)^{-1}$$

for a junction area of ~ 80 nm x 30 nm.

## 3. *WRITE* UNIT

The *Write* unit is essentially the same as that described in [5] and is based on the GSHE recently observed in high spin-orbit materials like platinum, tantalum and tungsten [5,8-10] or alloys like CuBi [11]. A charge current I gives rise to a spin current $I_s$ that carries $\hat{z}$ spins in the $\hat{y}$ direction

$$\vec{I}_s = \beta I \hat{z} \quad (2a)$$
$$where \; \beta = Spin-Hall \, Angle \times (A_s / A) \quad (2b)$$

$A_s$ and $A$ being the cross-sectional areas for the spin current and charge currents respectively. For a magnet with L= 80 nm this ratio could be ~ 40 if the thickness $t_0$ of the high spin-orbit metal layer is 2 nm. Based on the demonstrated spin-Hall angle of 0.3 in tungsten [9], this would give a charge to spin amplification factor of $\beta = 12$. Note, however, that the high resistivity of thin tungsten layers [9] could make the resistance "r" fairly large. Other materials like CuBi with comparable spin-Hall angles but lower resistivity [11] may be preferable.

We will now describe a possible scheme for combining a *Read* unit with a *Write* unit to form a spin switch that is *concatenable*, whereby the output of one switch can drive the next switch.

## 4. CONCATENABLE SPIN SWITCH

Our proposed spin switch (Fig.3a) consists of a *Write* unit and a *Read* unit that are *electrically isolated,* but *magnetically coupled* through a dipolar interaction that is strong enough to ensure that the angle $\theta$ between them always has a fixed value. The interaction could be ferromagnetic making $\theta = 0$, or antiferromagnetic making $\theta = \pi$ or even something in-between. Although the *Write* and *Read* units are shown side by side as is common in nanomagnetic logic [12], in practice it may be better to lay the *Read* unit on top of the *Write* unit to enhance the magnetic interaction. For our simulations we have assumed identical magnets each of which creates a dipolar field equal to the coercive field $H_K$ at the other magnet in a direction that keeps the two magnets anti-parallel.

Fig.3c shows the equivalent circuit for the spin switch. The input circuit determines the input current $V_{in}/(R_{in}+r)$ which through Eq.(2) determines the spin current entering the input magnet $\hat{m}'$. The output circuit describes the output voltage $V_{out}$ which is determined by the output magnet $\hat{m}$ and the associated fixed magnets $\pm \hat{M}$ through Eq.(1).

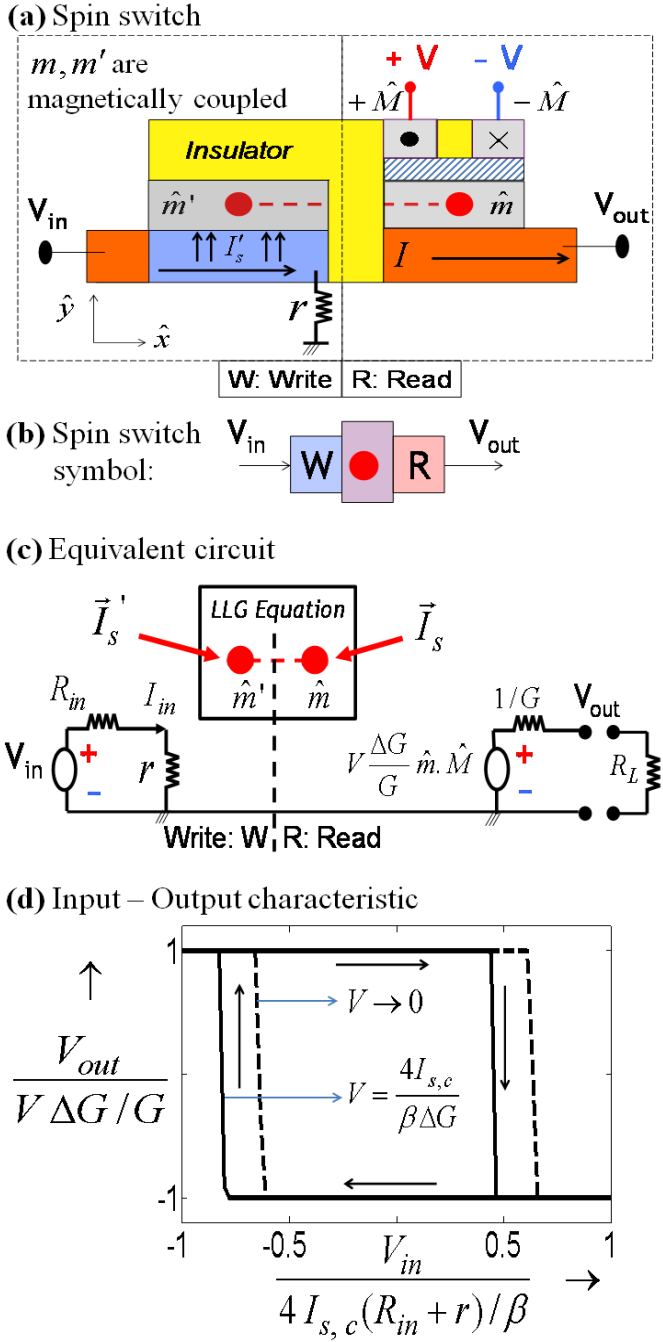

Figure 3 – (a) Proposed spin switch combining the *Write* and *Read* units from Fig.1b, with the magnet $\hat{m}$ from the *Read* unit magnetically coupled to the magnet $\hat{m}'$ from the *Write* unit, but electrically isolated from it. In practice it may be better to place one unit on top of the other rather than side-by-side to enhance the magnetic interaction. (b) Symbol representing the spin switch, showing a *Write* input and a *Read* output, with an inbuilt memory comprising a pair of coupled magnets represented by the red dot. (c) Equivalent circuit for spin switch. (d) Calculated hysteretic inverter-like characteristics for the spin switch with the output open-circuited ($R_L \to \infty$) using the model described in Appendices A,B.

We will assume that $\hat{m}$ (like $\hat{m}'$) has its easy axis along $\hat{Z}$, and the two fixed magnets $+\hat{M}$ and $-\hat{M}$ point along $+\hat{Z}$ and $-\hat{Z}$ respectively, so that

$$V_{out} = +V\Delta G/G \quad or \quad -V\Delta G/G \qquad (3)$$

depending on whether $\hat{m}$ points along $+\hat{M}$ or $-\hat{M}$. The easy axes $\hat{Z}$ need not be exactly aligned with the direction $\hat{z}$ of the spins injected by the *Write* unit (Eq.(2)). In our simulations we assume a small angle (~ 0.01 radian) between them.

To see why the spin switch should give rise to the input-output characteristics in Fig.3d, we note that if the current entering the write unit is large enough to generate a spin current $\beta V_{in}/(R_{in}+r)$ through the SHE that exceeds a certain critical value, it will switch the magnet $\hat{m}'$ to the $+\hat{Z}$ direction, putting the other magnet $\hat{m}$ of the pair in the $-\hat{Z}$ direction, so that the output voltage $V_{out} = -V\Delta G/G$ (see Eq.(3)). If we now reverse the input voltage beyond the critical value, the magnets are switched in the opposite direction with a reversal of the output voltage, resulting in a hysteretic inverter-like characteristic as shown in Fig.3d. Note that the sign of $V_{out}/V_{in}$ in Fig.3d could be changed by reversing either the sign of the $\beta$ or the V associated with the *Write* and *Read* units respectively.

In Fig.3d we have normalized the input spin current $\beta V_{in}/(R_{in}+r)$ by $4I_{s,c}$, $I_{s,c}$ being the critical current for one of the magnets ($\hat{m}'$ or $\hat{m}$ which are assumed identical) [5]

$$I_{s,c} = (2q/\hbar)\alpha\mu_0 M_s \Omega(H_K + M_s/2) \qquad (4)$$

where $M_s$ is the saturation magnetization, $H_K$ is the coercive field, $\alpha$ is the damping parameter and $\Omega$ is the volume of the magnet. As we might expect, a spin current ~2 $I_{s,c}$ is needed to switch since two magnets are coupled together. We consider only in-plane magnets, leaving perpendicular magnetization for future work.

Note that with negligible voltage applied to the *Read* unit, the switching current is the same for -Z to +Z and for +Z to -Z. But a voltage V applied to the *Read* unit shifts the loop along the x-axis because it injects a spin current along –Z which aids the *Write* unit when the *Read* unit is switching to –Z. The value of V used for the solid curve in Fig.3d is chosen to exceed what is needed to ensure that a *Read* unit generates an input voltage of $V_{in} = V\Delta G/G$ for the next *Write* unit that is large enough to switch it assuming $r << R_{in} = 1/G$. According to Fig.3d this requires

$$V \geq \frac{4I_{s,c}}{\beta \Delta G}(1+Gr) \qquad (5)$$

With $\Omega = 80nm \times 100nm \times 1.6nm$, and $\mu_0 M_s \sim 1T$, $\mu_0 H_K \sim .02T$, $\alpha \sim .01$ we have $I_{s,c} \sim 160\mu A$. Using $\beta = 12$, $Gr<<1$, $\Delta G/G \sim 0.4$, $G^{-1} \sim 1.1\ K-ohms$ from our earlier estimates, we have $V \sim 150\ mV$. We use these parameters for all simulations, though they have not been optimized and other choices may be preferred depending on the application. Details of the simulation model are described in Appendices A and B.

## 5. RING OSCILLATOR

Fig.4 shows an example of how spin switches of the type shown in Fig.3a (represented by the symbol in Fig.3b), are interconnected to implement a circuit function, in this case a ring oscillator. If the voltages V on the *Read* units exceed the threshold value given by Eq.(5), then each unit will try to switch the corresponding magnet of the next unit anti-parallel to itself. But with an odd number of magnets (three in this case) in the loop, there is no way to make all neighboring magnets anti-parallel and so no satisfactory steady state can be achieved. Instead each magnet switches the magnet downstream which in turn switches the next magnet, resulting in continuous oscillations as shown in Fig.4 from our simulations.

Such oscillations are well-known in the context of CMOS inverters and Fig.4 shows that our proposed spin switch has these transistor-like characteristics allowing it to generate an oscillatory output from a dc power supply without any external amplifier or clock. This requires properties like directivity and input-output isolation needed for SCL units. We have also simulated more sophisticated circuits involving majority gates with multiple fan-out (see Appendix B), suggesting that the proposed spin switch could be used to construct large scale circuits.

## 6. RECONFIGURABLE COMPARATOR

Figure 5 shows a different device that could be implemented by interconnecting our proposed spin switches (Fig.3a). It should provide an output that correlates the incoming signal $\{X_n\}$ with a reconfigurable reference signal $\{Y_n\}$ stored in the $m_z$ of the switches that could be any string of +1's and -1's of length N, N being a large number.

Since the output current (see Eq.(1)) of each Read unit is a product of V (~ $X_n$) and $m_z$ (~ $Y_n$) it is determined by $X_n Y_n$ which are all added up to drive the output magnet. If the sequence $\{X\}$ is an exact match to $\{Y\}$, then the output voltage will be N, since every $X_n*Y_n$ will equal +1, being either (+1)*(+1) or (-1)*(-1). If $\{X\}$ matches $\{Y\}$ in (N-n) instances with n mismatches, the output will be N-(2*n) since every mismatch lowers output by 2. If we set the threshold for the output magnet to N-(2*$N_e$) then the output will respond for all $\{X\}$ that matches the reference $\{Y\}$ within a tolerance of $N_e$ errors. The inset in Fig.5 shows an example with $N_e$=0.

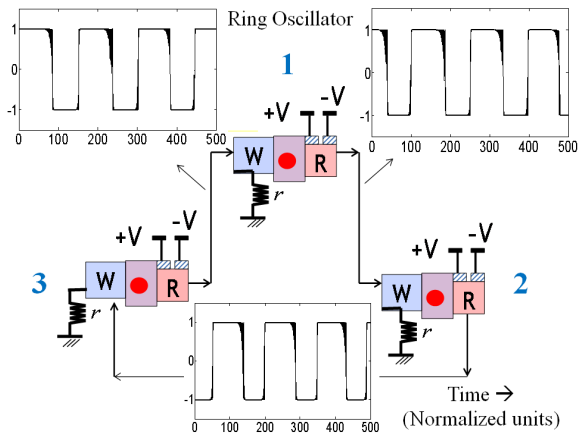

Figure 4 – A ring oscillator implemented by interconnecting three spin switches A, B and C of the type shown in Fig.3a. The plots show the currents normalized to $4I_{s,c}/\beta$, in each of the three connecting wires.

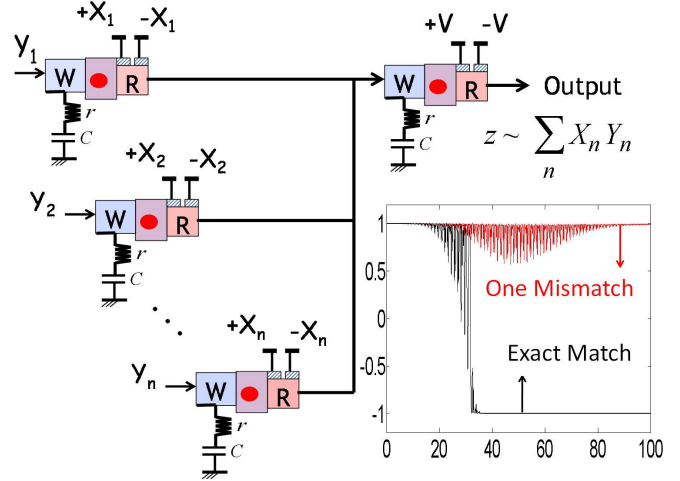

Figure 5 – An example of a device that could be implemented by interconnecting our proposed spin switches (Fig.3a) which should provide an output that correlates the incoming signal $\{X_n\}$ with a reconfigurable reference signal $\{Y_n\}$ stored in the switches. Inset shows response of output magnetization as a function of time (normalized). Threshold is adjusted such that the magnet is switched only if all 20 bits of $\{Y\}$ match all 20 bits of $\{X\}$. With even one mismatch the output fails to switch. Note that no middle circuitry for signal conversion or amplification is involved.

## 7. WEIGHTED INTERCONNECTIONS

An important distinction of the spin switch with a CMOS device is the representation and storage of information in nanomagnets rather than in voltages. This makes the switch non-volatile, and also allows us to use the voltages to change the weight of the connection between a *Read* and a *Write* unit in an analog manner, with magnets naturally providing the digitization. We could extend Eq.(5) to define the *weight* $a_{ij}$ as the ratio of the current from the *Read* magnet of unit "j" normalized to the critical current needed to switch the *Write* magnet of unit "i":

$$a_{i,j} = \frac{(V\Delta G)_j}{(4I_{s,c}/\beta)_i} \frac{1}{1+G_{tot}\, r_i} \qquad (6)$$

where $G_{tot}$ is the sum of the $G_j$ for all *Read* units connected to the same Write unit.

This feature could make it possible to interconnect our proposed spin switches to implement hardware neural networks [13] where both the sign and the magnitude of the weights $a_{ij}$ (Eq.(6)) associated with each *Read-Write* connection can be conveniently adjusted through the voltages V applied to the *Read* unit and the gain $\beta$ of the *Write* unit. Using present day technology it should be possible to implement weights that exceed switching thresholds either individually (Fig.3d) or collectively (Fig.5), but even subthreshold networks could find use in probabilistic logic as discussed for ASL in [14]. Note also the additional factor $(1+G_{tot}r_i)$ that enters the expression for the weights $a_{ij}$. Ordinarily the resistances $r_i$ arise from that of the high spin-orbit material in the *Write* units. But it may be possible to add a phase change resistance in series that could provide an automatic adjustment of weights, perhaps making it possible for networks to "learn" [15].

## 8. SUMMARY

We have shown that the established physics of SVs/MTJ's could be used to construct a *Read* unit that converts the information stored in a nanomagnet into a current which can be used to operate a *Write* unit based on the physics of the GSHE. We then show how the *Write* and *Read* units could be integrated into a single spin switch showing inverter like characteristics having input-output isolation, directivity and fan-out similar to CMOS inverters, but with the information stored in nanomagnets making it non-volatile.

The gain $\beta$ made possible by the GSHE allows us to engineer a *Write-Read* asymmetry whereby the switch follows the *Write* signal and not the *Read* signal. This is a key requirement for SCL units analogous to the well-known property of field effect transistors (FET's) that makes them switch in response to a gate voltage (which "writes" the input onto the state of the transistor) but not in response to a comparable drain voltage (which "reads" the state of the transistor). We have analyzed only the simplest design with identical *Read* and *Write* magnets leaving it to future work to determine if the *Write-Read* asymmetry can be enhanced through other means such as making one magnet softer or smaller than the other. Indeed even the basic *Read* and *Write* mechanisms could possibly be improved for lower switching energy, delay and standby power.

However, the key point is that the proposed spin switches can be interconnected just with passive circuit elements, to perform logic operations, without any external circuitry for signal conversion or amplification. Moreover, since the digitization and storage occurs in the magnets, the switches are non-volatile and reconfigurable and the voltages can be used to implement analog "weighting" for non-Boolean logic.

### ACKNOWLEDGEMENTS

It is a pleasure to thank Vinh Quang Diep for carefully going over the manuscript and, along with Seokmin Hong, for helpful discussions regarding the giant spin Hall effect. S.D. acknowledges support from the Institute for Nanoelectronics Discovery and Exploration (INDEX).

**Appendix A: Modeling a spin switch**

Each spin switch can be modeled using a pair of LLG equations to model the pair of magnets as two macrospins $\hat{m}'$ and $\hat{m}$ coupled by the dipolar interaction.

$$(1+\alpha'^2)\frac{d\hat{m}'}{dt} = -|\gamma|\mu_0 \hat{m}' \times \vec{H}' - \alpha'|\gamma|\mu_0 \hat{m}' \times \hat{m}' \times \vec{H}'$$
$$- \hat{m}' \times \hat{m}' \times \frac{\vec{I}_s'}{qN_s'} + \alpha' \hat{m}' \times \frac{\vec{I}_s'}{qN_s'}$$
(A1a)

$$(1+\alpha^2)\frac{d\hat{m}}{dt} = -|\gamma|\mu_0 \hat{m} \times \vec{H} - \alpha|\gamma|\mu_0 \hat{m} \times \hat{m} \times \vec{H}$$
$$- \hat{m} \times \hat{m} \times \frac{\vec{I}_s}{qN_s} + \alpha \hat{m} \times \frac{\vec{I}_s}{qN_s}$$
**(A1b)**

Here $\gamma$ is the gyromagnetic ratio, $\alpha, \alpha'$, the damping parameter and $N_s = M_s V / \mu_B$, $N_s' = (M_s V)' / \mu_B$ are the number of spins comprising each magnet ($\mu_B$: Bohr magneton).

It is convenient to transform Eqs.(A1a,b) to dimensionless variables

$$\frac{1+\alpha'^2}{1+\alpha^2}\frac{d\hat{m}'}{d\tau} = -\hat{m}' \times \vec{h}' - \alpha' \hat{m}' \times \hat{m}' \times \vec{h}'$$
$$- \hat{m}' \times \hat{m}' \times \vec{i}_s' + \alpha' \hat{m}' \times \vec{i}_s'$$
(A2a)

$$\frac{d\hat{m}}{d\tau} = -\hat{m} \times \vec{h} - \alpha \hat{m} \times \hat{m} \times \vec{h}$$
$$- \hat{m} \times \hat{m} \times \vec{i}_s + \alpha \hat{m} \times \vec{i}_s$$
(A2b)

where $\tau \equiv \dfrac{|\gamma|\mu_0 H_K t}{1+\alpha^2}$

$\vec{h} = \vec{H}/H_K, \vec{h}' = \vec{H}'/H_K$

(A3)

and $\vec{i}_s = \dfrac{\vec{I}_s}{qN_s|\gamma|\mu_0 H_K} = \dfrac{\vec{I}_s}{(2q/\hbar)\mu_0 H_K M_s \Omega}$

$\vec{i}_s' = \dfrac{\vec{I}_s'}{(2q/\hbar)\mu_0 H_K M_s \Omega}$

The total field entering Eqs.(A1a,b)

$$\vec{H} = H_k m_Z \hat{Z} - H_d m_y \hat{y} - H_f \hat{m}'$$ (A4a)
$$\vec{H}' = H_k' m_Z' \hat{Z} - H_d' m_y' \hat{y} - H_b \hat{m}$$ (A4b)

includes the easy axis fields ($H_k, H_k'$), the demagnetizing fields ($H_d, H_d'$) plus the dipolar fields ($H_f, H_b$).

An exact treatment of the dipolar fields would require a detailed consideration of the shape of each magnet [A1], but for our purpose the approximate expression in Eqs.(A4) should be adequate with $H_f = (M_s A_s)'/d^2$, $H_b = (M_s A_s)/d^2$, $M_s, M_s'$ being the saturation magnetizations, $A_s, A_s'$, the areas (in x-z plane) and d, the distance (along x) between the magnets.

Since we want the *Write* magnet $\hat{m}'$ to switch the *Read* magnet $\hat{m}$, it may help speed up the process if we design the forward interaction $H_f$ to be larger than the backward one $H_b$. The simplest way to achieve this is to make the *Write* magnet larger than the *Read* magnet, but more sophisticated approaches based on engineering material parameters may be possible too.

For our examples in this paper we have used identical *Read* and *Write* magnets with the following parameters:

$$\alpha = .01, H_d = 50 H_k, H_f = H_b = H_k = 0.02T/\mu_0$$
$$\mu_0 M_s = 1T, \Omega = 80nm \times 100nm \times 1.6nm$$
$$\rightarrow N_s \sim 1.1 \times 10^6$$

so that from Eq.(4)
$$I_{s,c} \sim 160 \ \mu A$$

The time axis in all plots in this paper represents the dimensionless time defined in Eq.(A3):

$$\tau \sim \frac{t}{285 \ ps}$$

How the spin currents $\vec{i}_s'$ and $\vec{i}_s$ into the *Write* and *Read* units are calculated is discussed in Appendix B (Eqs. (B3) and (B5)).

Reference

[A1] See for example, E.Y. Tsymbal, "Theory of Magnetostatic Coupling in Thin-Film Rectangular Magnetic Elements," *Appl. Phys. Lett.* **77**, 2740 (2000).

**Appendix B: Modeling spin switch networks**

In this Appendix we will illustrate our model with a concrete simulation illustrating the operation of a multistage Boolean gate constructed using the basic spin switch that we discussed. Consider a majority gate with three *Read* units driving three input switches 1, 2 and 3 respectively which drive switch 4 which in turn drives two output switches 5 and 6 (Fig.B1). For generality, we have added a capacitor C in series with the resistor r in every *Write* unit.

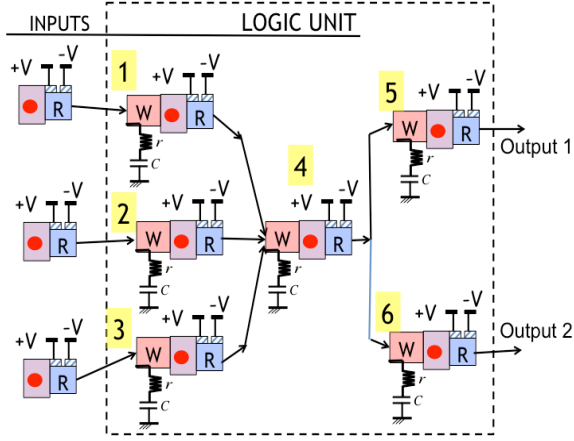

Figure B1 – A majority gate with three Read units driving three input devices 1, 2 and 3 respectively which drive device 4 which in turn drives two output devices 5 and 6.

Since this gate includes six spin switches each having a *Write* and a *Read* unit, we need to model the dynamics of twelve magnetization vectors $\hat{m}_{1W}, \hat{m}_{1R}, \cdots, \hat{m}_{6W}, \hat{m}_{6R}$ described by LLG equations as described in Appendix A (Eq.(A1)). We need an expression for the spin currents driving the different *Write* units which are derived from the *Read* units of the previous stage. In addition there are spin currents driving the *Read* units arising from the fixed magnets $\pm \hat{M} = \pm \hat{Z}$. These unwanted *Read* spin currents affect the switching of the magnets (see solid and dashed curves in Fig.3d), and the $\beta$ associated with the SHE helps keep these small compared to the *Write* spin currents.

Both the *Read* and *Write* spin currents can be determined using the equivalent circuit for a single *Read* unit driving a single *Write* unit (Fig.2), redrawn here with the additional capacitor C (Fig.B2). For generality we have also assumed that there are $n_i$ identical *Write* units driving $n_o$ identical *Read* units. For example, in the majority gate shown in Fig.B1, (a) unit 4W is driven by 1R, 2R and 3R making $n_i=3$, $n_o=1$, while (b) units 5W and 6W are driven by 4R making $n_i=1$, $n_o=2$.

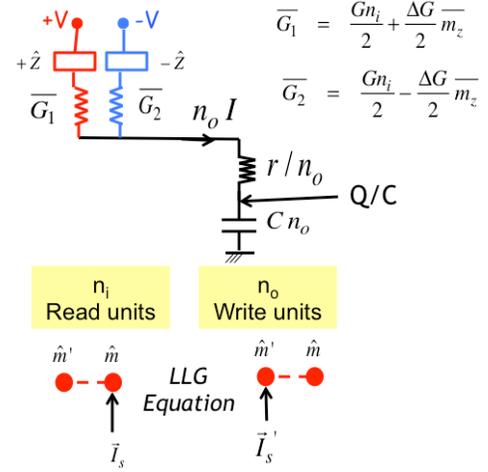

Figure B2 – Equivalent circuit for $n_i$ *Read* units driving $n_o$ *Write* units. Compared to Fig.2 in the text we have also include a series capacitor, C. The overbars on the conductances and the magnetization represents a sum over all the inputs.

The total current flowing into the Write unit is the sum of the currents flowing through $\overline{G_1}$ and $\overline{G_2}$:

$$n_o I(t) = \overline{G_1}(t)\left(V - I(t)r - \frac{Q(t)}{C}\right)$$
$$+ \overline{G_2}(t)\left(-V - I(t)r - \frac{Q(t)}{C}\right)$$
$$= V\Delta G \overline{m_z} - \left(Ir + \frac{Q}{C}\right)n_i G$$

$$I = \frac{dQ}{dt} = \frac{V\Delta G \overline{m_z} - n_i QG/C}{n_o + n_i Gr}$$

$$= \frac{n_i G}{n_o + n_i Gr}\left(VX - \frac{Q}{C}\right) \quad (B1)$$

where $\quad X \equiv \Delta G \overline{m_z}/n_i G \quad (B2)$

The twelve LLG equations for $d\hat{m}/dt$ in Eq.(A1) are augmented with six equations for dQ/dt like Eq.(B1) for each of the six capacitors.

The *Write* spin current equals the amplification factor $\beta$ times the charge current in Eq.(B1):

$$\vec{I}_s' = \hat{z}\beta I = \hat{z}\beta \frac{n_i G}{n_o + n_i Gr}\left(VX - \frac{Q}{C}\right) \quad (B3)$$

If the capacitors are absent we have the same result, but with Q/C in Eq.(B3) set equal to zero.

The spin current at a particular *Read* unit is obtained from the difference in the currents flowing through $G_1$ and $G_2$ of a single *Read* unit in Fig.B2:

$$\vec{I}_s = \hat{Z}\left[G_1\left(V - Ir - \frac{Q}{C}\right) - G_2\left(-V - Ir - \frac{Q}{C}\right)\right]$$

$$= \hat{Z}\left[VG - \Delta G m_z\left(Ir + \frac{Q}{C}\right)\right]$$

$$= \hat{Z}\ G\left[V - X_1\left(Ir + \frac{Q}{C}\right)\right]$$

where $\quad X_1 \equiv \Delta G m_z / G \qquad$ (B4)

Using Eq.(B1) we can write

$$Ir + \frac{Q}{C} = \frac{n_i Gr}{n_o + n_i Gr}VX + \frac{Q}{C}\frac{n_o}{n_o + n_i Gr}$$

so that from Eq.(B3)

$$\vec{I}_s = \hat{Z}\left[VG\left(1 - \frac{X_1 X n_i Gr}{n_o + n_i Gr}\right) - \frac{QG}{C}\left(\frac{n_o X_1}{n_o + n_i Gr}\right)\right] \quad (B5)$$

Following are some results obtained from the solution of 42 coupled first-order differential equations, 36 for the three components of the twelve magnetization vectors, and 6 for the charge on each capacitor. The magnet parameters are those listed at the end of Appendix A, while the circuit parameters are those discussed in the paper (like $\Delta G/G = 0.4, \beta = 12, Gr \ll 1$) with $C/G = 500$. The same parameters were used for Fig.3d and Fig.4 except that the series capacitor was not included, so that the Q/C terms in Eqs.(B3) and (B5) are absent and Eqs.(B1) for the dQ/dt's are not needed. All our simulations assume a small angle (0.01 radian) between $\tilde{z}$ (see Eq.(B4)) and $\hat{Z}$ (see Eq.(B5)).

All W magnets were initialized in the -1 state while the R magnets were initialized in the +1 state. The output voltages are given by $V \Delta G / G$ times the $m_z(t)$ for magnets 5R and 6R shown in Fig.B3. A voltage of $V = 10 I_{s,c} / \beta \Delta G$ was used for the simulation. The three inputs were assumed to be +1, -1 and +1, causing both W and R magnets for 1 and 3 to change their states, while magnets 2W and 2R remain in their initial state (Fig.B4). Initially 4W switches to +1 (and 4R to -1) since 1R, 2R and 3R are all initialized to +1. But once 1R and are switched to -1, 4W follows the majority and switches to -1, making 4R switch to +1. 5W and 6W then follow 4R and switch to +1, making their dipole coupled partners 5R and 6R switch to -1.

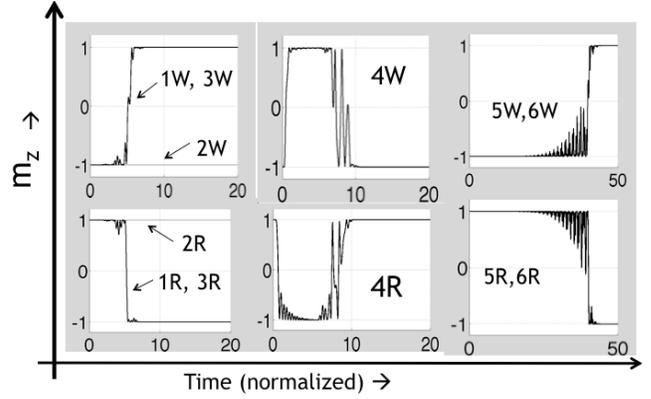

Figure B3 – Z-components of the magnetization vectors of the six Write (W) and the six Read (R) magnets. All W magnets were initialized in the -1 state while the Read magnets were initialized in the +1 state. The three inputs were assumed to be +1, -1 and +1, causing both W and R magnets for 1 and 3 to change their states, while magnets 2W and 2R remain in their initial state.

Fig.B4 shows the currents into each of the Write units which look constant on a short time scale, but decay when we look on a time scale comparable to the charging time constant $C/G$. This could be useful in reducing standby power though the details need careful consideration beyond the scope of this paper.

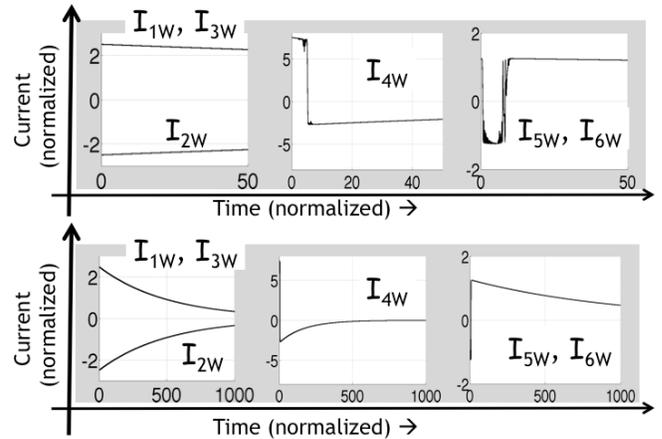

Figure B4 – Calculated currents (normalized to $4 I_{s,c} / \beta$) into each of the write units shown both on a short time scale and on a long time scale compared to $C/G = 500$ in the normalized time units defined in Appendix A.